\documentclass[final,5p,nopreprintline,times,twocolumn]{elsarticle}
\biboptions{sort&compress}
\journal{Physics Letters B}

\usepackage{amsmath,slashed}
\usepackage{graphicx, graphics, color}
\usepackage{dcolumn}
\usepackage{xspace}
\usepackage{enumerate}
\usepackage{varwidth}
\usepackage{microtype}

\usepackage{balance}

\usepackage{physics}
\usepackage{amssymb}
\usepackage{mathtools}
\usepackage{dsfont}
\usepackage{multirow}
\usepackage{multicol}
\usepackage{float}

\usepackage{booktabs}
\usepackage[flushleft]{threeparttable}

\usepackage[hyperfootnotes=true,colorlinks=false,citecolor=blue,linkcolor=blue,urlcolor=blue]{hyperref}

\usepackage{orcidlink}

\makeatletter
\newcounter{savesection}
\newcounter{apdxsection}
\renewcommand\appendix{\par
  \setcounter{savesection}{\value{section}}%
  \setcounter{section}{\value{apdxsection}}%
  \setcounter{subsection}{0}%
  \gdef\thesection{\@Alph\c@section}}
\makeatother

\newcommand{\MeV}{\,\text{MeV}}

\newcommand{\Br}{\text{Br}}

\newcommand{\beq}{\begin{equation}}
\newcommand{\eeq}{\end{equation}}

\newcommand{\elem}[2]{\ensuremath{^{#2}\text{#1}}}

\newcommand{\fmi}{\ensuremath{\text{fm}^{-1}}}

\newcommand{\nnlo}{\ensuremath{\text{N$^2$LO}}}
\newcommand{\nnnlo}{\ensuremath{\text{N$^3$LO}}}
\newcommand{\hebelerA}{\ensuremath{\text{1.8/2.0~(EM)}}}
\newcommand{\hebelerB}{\ensuremath{\text{2.0/2.0~(EM)}}}
\newcommand{\hebelerC}{\ensuremath{\text{2.2/2.0~(EM)}}}
\newcommand{\hebelerD}{\ensuremath{\text{2.0/2.0~(PWA)}}}
\newcommand{\nnlosat}{\ensuremath{\text{NNLO}_\text{sat}}}
\newcommand{\dnnlogo}{\ensuremath{\Delta\text{NNLO}_\text{GO}}}
\newcommand{\arthuisA}{\ensuremath{\text{1.8/2.0~(EM7.5)}}}
\newcommand{\arthuisB}{\ensuremath{\text{1.8/2.0~(sim7.5)}}}

\newcommand{\Id}{\mathds{1}}

\newcommand{\rr}{\mathbf{r}}
\newcommand{\rrhat}{\hat{\mathbf{r}}}

\allowdisplaybreaks

\begin{document}

\title{Ab initio calculations of overlap integrals for $\mu\to e$ conversion in nuclei}

\author[tud,emmi,mpik,olcf,ornl]{Matthias Heinz\orcidlink{0000-0002-6363-0056}}
\author[bern]{Martin Hoferichter\orcidlink{0000-0003-1113-9377}}
\author[tsuk,tud,emmi,mpik]{Takayuki Miyagi\orcidlink{0000-0002-6529-4164}}
\author[bern]{Frederic Noël\orcidlink{0000-0002-7450-7213}}
\author[tud,emmi,mpik]{Achim Schwenk\orcidlink{0000-0001-8027-4076}}

\address[tud]{Technische Universit\"at Darmstadt, Department of Physics, 64289 Darmstadt, Germany}
\address[emmi]{ExtreMe Matter Institute EMMI, GSI Helmholtzzentrum f\"ur Schwerionenforschung GmbH, 64291 Darmstadt, Germany}
\address[mpik]{Max-Planck-Institut f\"ur Kernphysik, Saupfercheckweg 1, 69117 Heidelberg, Germany}
\address[olcf]{National Center for Computational Sciences, Oak Ridge National Laboratory, Oak Ridge, TN 37831, USA}
\address[ornl]{Physics Division, Oak Ridge National Laboratory, Oak Ridge, TN 37831, USA}
\address[bern]{Albert Einstein Center for Fundamental Physics, Institute for Theoretical Physics, University of Bern, Sidlerstrasse 5, 3012 Bern, Switzerland}
\address[tsuk]{Center for Computational Sciences, University of Tsukuba, 1-1-1 Tennodai, Tsukuba 305-8577, Japan}

\begin{abstract}
 The rate for $\mu\to e$ conversion in nuclei is set to provide the most stringent test of lepton-flavor symmetry and a window into physics beyond the Standard Model. However, to disentangle new lepton-flavor-violating interactions, in combination with information from $\mu\to e\gamma$ and $\mu\to 3e$, it is critical that uncertainties at each step of the analysis be controlled and fully quantified. In this regard, nuclear response functions related to the coupling to neutrons are notoriously problematic, since they are not directly constrained by experiment. We address these shortcomings by combining ab initio calculations with a recently improved determination of charge distributions from electron scattering by exploiting strong correlations among charge, point-proton, and point-neutron radii and densities. We present overlap integrals for $^{27}$Al, $^{48}$Ca, and $^{48}$Ti including full covariance matrices, allowing, for the first time, for a comprehensive consideration of nuclear structure uncertainties in the interpretation of $\mu\to e$ experiments.
\end{abstract}

\begin{keyword}
$\mu \to e$ conversion in nuclei, ab initio calculations

\end{keyword}

\maketitle

\section{Introduction}

One of the frontiers of contemporary particle physics is the search for physics beyond the Standard Model (BSM) in low-energy precision observables. Among the most promising probes are observables that test so-called accidental symmetries of the SM, such as the conservation of lepton flavor or baryon number, as these are not based on fundamental principles, and BSM extensions may well violate these symmetries. While the mere observation of these processes would thus already constitute a BSM signal, a rigorous theoretical description is critical to be able to draw conclusions on possible BSM scenarios, and in many cases the most critical aspects concern the hadronic and nuclear matrix elements, forming the bridge between the fundamental degrees of freedom and the actual observables.    

Lepton-flavor-violating (LFV) processes are particularly attractive, since they are forbidden in the SM apart from tiny corrections due to neutrino oscillations, yielding rates at the level of ${\mathcal O}(10^{-50})$. The current leading limits on the purely leptonic channels are on muon decays to electrons, with $\Br[\mu\to e\gamma]<4.2\times 10^{-13}$~\cite{MEG:2016leq} and $\Br[\mu\to 3e]<1.0\times 10^{-12}$~\cite{SINDRUM:1987nra} (here and below at $90\,\%$ confidence level). These limits are set to improve at MEG II~\cite{MEGII:2018kmf} and  Mu3e~\cite{Mu3e:2020gyw}, respectively (and potentially beyond~\cite{Aiba:2021bxe}). In this Letter, we consider the process 
$\mu\to e$ conversion in nuclei, in which a muon bound in an atom converts into an electron in the Coulomb field of the nucleus, and then is ejected with the energy converted from the muon mass. 
The current best limits are given by the SINDRUM-II experiment as
\begin{align}
 \Br[\mu\to e, \text{Ti}] &< 6.1 \times 10^{-13},\nonumber\\
 \Br[\mu\to e, \text{Au}] &< 7 \times 10^{-13},
 \label{muelimits}
\end{align}
measured on titanium~\cite{Wintz:1998rp} and gold targets~\cite{SINDRUMII:2006dvw}, respectively.\footnote{Conventionally, these limits are normalized to muon capture~\cite{Suzuki:1987jf}.  The earlier limit $\Br[\mu\to e, \text{Ti}]<4.3\times 10^{-12}$~\cite{SINDRUMII:1993gxf} is superseded by Ref.~\cite{Wintz:1998rp} as the final result from the SINDRUM-II experiment.}
These limits are set to improve by up to four orders of magnitude at the upcoming experiments Mu2e~\cite{Mu2e:2014fns} and COMET~\cite{COMET:2018auw} using aluminum targets, making this process the most stringent test of LFV to date, especially for BSM scenarios in which the LFV is mediated by interactions involving quark degrees of freedom~\cite{Mihara:2013zna}. These prospects strongly motivate the development of a robust theoretical description of $\mu \to e$ conversion in nuclei. Given the vastly different scales in the problem, ranging from the BSM scale down to the nuclear scale, this is most efficiently achieved using effective field theory (EFT).

\section{EFT approach to $\mu\to e$ conversion in nuclei}

In EFT, $\mu\to e$ conversion in nuclei is described in terms of effective LFV operators defined at the BSM scale. One must then account for all the different scales that play a role as these operators are evolved down to the nuclear scale~\cite{Kitano:2002mt,Cirigliano:2009bz,Petrov:2013vka,Crivellin:2013hpa,Crivellin:2014cta,Davidson:2018kud,Rule:2021oxe,Cirigliano:2022ekw,Hoferichter:2022mna,Haxton:2022piv,Borrel:2024ylg,Noel:2024led,Noel:2024swe,Haxton:2024lyc,Delzanno:2024ooj}.
To this end, first renormalization group corrections need to be considered for the evolution to the lower scales~\cite{Crivellin:2017rmk,Cirigliano:2017azj,Davidson:2017nrp}. Then hadronic matrix elements turn the quark-level interactions into hadronic ones, and nuclear matrix elements account for the strong-interaction effects of embedding the nucleons into the atomic nucleus. Finally, Coulomb corrections need to be considered, which characterize the influence of the potential of the nucleus on the initial bound-state muon and the ejected electron. Using such a framework and combining it with complementary information from $\mu \to e \gamma$ and $\mu \to 3 e$, it becomes possible to disentangle different underlying sources of LFV~\cite{Davidson:2020hkf,Davidson:2022nnl,Ardu:2023yyw,Ardu:2024bua}.

Robustly drawing conclusions regarding the underlying LFV interactions from such limits requires uncertainties to be controlled and quantified at each step of the theoretical description, notably also for the hadronic and nuclear matrix elements and the Coulomb corrections. The nuclear response functions are particularly intricate, especially those related to the couplings to neutrons as they are not directly constrained by experiment. Only recently, direct insights into neutron densities of a few selected nuclei became available via parity-violating electron scattering (PVES)~\cite{PREX:2021umo,Qweak:2021ijt,CREX:2022kgg}. In this Letter, we address the shortcomings of previously unquantified uncertainties for the nuclear structure input by predicting and exploiting strong correlations among charge, proton, and neutron matrix elements from ab initio calculations using the in-medium similarity renormalization group (IMSRG)~\cite{Tsukiyama:2010rj, Hergert:2015awm,Stroberg:2016ung,Stroberg:2019mxo,Stroberg:2019bch} with state-of-the-art interactions from chiral EFT~\cite{Hebeler:2010xb,Ekstrom:2015rta,Jiang:2020the,Hu:2021trw,Arthuis:2024mnl}.

The leading contributions to $\mu\to e$ conversion originate from scalar, vector, and dipole interactions, which couple in a spin-independent (SI) way to the nucleus and thus show a coherent enhancement with the number of nucleons in the nucleus. The SI $\mu \to e$ conversion rate is conventionally expressed in terms of so-called overlap integrals~\cite{Kitano:2002mt}, labeled as $S^{(N)}$, $V^{(N)}$, and $D$, with $N=n,p$,
according to
\begin{align}
    \Br^\text{SI}_{\mu\to e}&=\frac{4m_\mu^5}{\Gamma_\text{cap}}\sum_{Y=L,R}\bigg|\sum_{I_i} \bar C_Y^{I_i} ~ I_i ~ \bigg|^2,
    \label{eq:BR_SI}
\end{align}
with the muon capture rate $\Gamma_\text{cap}$ and where $I_{i}$ runs over all overlap integrals. These overlap integrals, which connect the underlying physics contained in the prefactors
$\bar{C}^{I_i}_Y$ (given as a combination of Wilson coefficients and hadronic matrix elements~\cite{Hoferichter:2022mna}, see Supplementary Material for explicit expressions) to the decay rate, are the central objects of this study.

Obtaining fully quantified ab initio uncertainties for these overlap integrals is
challenging, given the direct sensitivity to proton $\rho_p$, neutron $\rho_n$, or charge distributions $\rho_\text{ch}$ [see  Eq.~\eqref{eq:S,V,D_kitano} below]. Furthermore, due to Coulomb distortions of the lepton wave functions, uncertainties in the nuclear charge distribution $\rho_\text{ch}$ also propagate in an indirect way. The latter aspect was recently addressed by an improved extraction of charge distributions from elastic electron--nucleus scattering including statistical and systematic uncertainty estimates and correlations~\cite{Noel:2024led}. Using these results, quantified uncertainties for the dipole overlap integrals $D$ for the nuclei considered could already be provided, as $D$ is fully determined by the charge distribution. In this Letter, we now address the uncertainties originating from the proton and neutron distributions $\rho_p$ and $\rho_n$, necessary for $S^{(N)}$, $V^{(N)}$ and the correlations among the various overlap integrals.

We focus on the isotopes $^{27}$Al, to be used in the upcoming Mu2e and COMET experiments, 
and $^{48}$Ti, with  $73.72\,\%$ by far the most abundant titanium isotope and thus relevant for the previous SINDRUM-II experiment. 
In addition, we consider $^{48}$Ca, which is a valuable benchmark for nuclear structure calculations~\cite{Hagen:2015yea,Heinz:2024juw}, is relevant in the context of PVES~\cite{CREX:2022kgg}, and whose charge distribution was measured precisely in electron scattering experiments.

\section{Ab initio calculations}

Ab initio calculations of nuclei are now able to simulate systems as heavy as \elem{Pb}{208}~\cite{Miyagi:2021pdc, Hu:2021trw, Hebeler:2022aui, Arthuis:2024mnl, Door:2024qqz}, provide a global description of medium-mass nuclei including deformation~\cite{Stroberg:2019bch, Frosini:2021sxj, Hagen:2022tqp, Sun:2024iht}, and compute nuclear responses necessary for a microscopic description of fundamental interactions in nuclei~\cite{Gazda:2016mrp, Payne:2019wvy, Simonis:2019spj, Lovato:2020kba, Sobczyk:2021dwm, Hu:2021awl}.
We employ nuclear forces from chiral EFT rooted in quantum chromodynamics (QCD)~\cite{Epelbaum:2008ga, Machleidt:2011zz}.
Such forces are inherently uncertain due to truncations in the EFT, unknown short-range couplings that must be fit to data, and residual regularization scale and scheme dependence.
To systematically explore this uncertainty, we consider a large ensemble of Hamiltonians consisting of nucleon--nucleon (NN) and three-nucleon (3N) potentials that differ in their construction within chiral EFT, their regularization scale, and how they are fit to data.
These include Hamiltonians fit only to two-, three-, and four-nucleon systems~\cite{Hebeler:2010xb}, Hamiltonians additionally optimized to bulk properties of medium-light nuclei and nuclear matter~\cite{Ekstrom:2015rta, Jiang:2020the, Arthuis:2024mnl}, and an ensemble of Hamiltonians constructed using a history matching procedure based on two- through four-nucleon systems and \elem{O}{16}~\cite{Hu:2021trw}.

\begin{figure}[t!]%
    \centering
    \includegraphics[width=0.99\linewidth,trim= 0 10 0 5,clip]{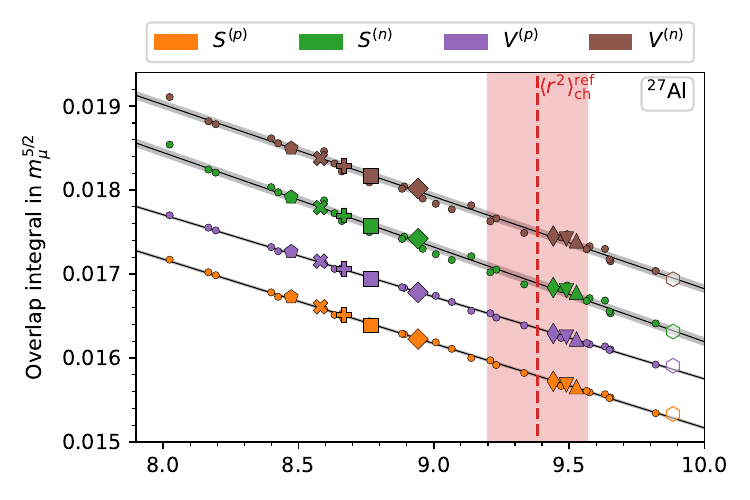}
    \includegraphics[width=0.99\linewidth,trim= 0 10 0 5,clip]{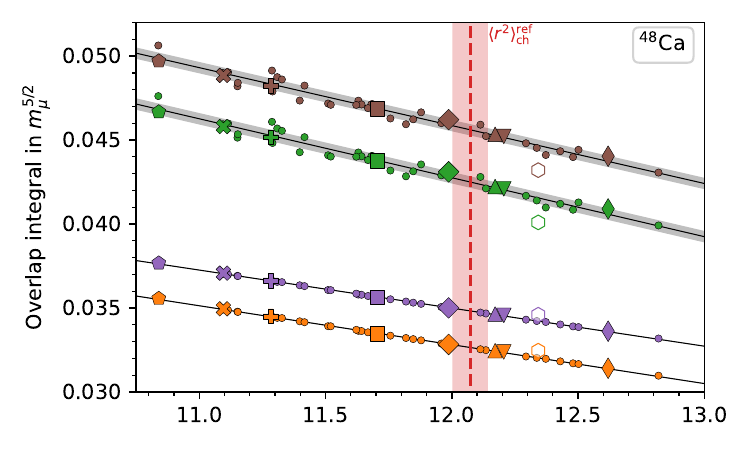}
    \includegraphics[width=0.99\linewidth,trim= 0 10 0 5,clip]{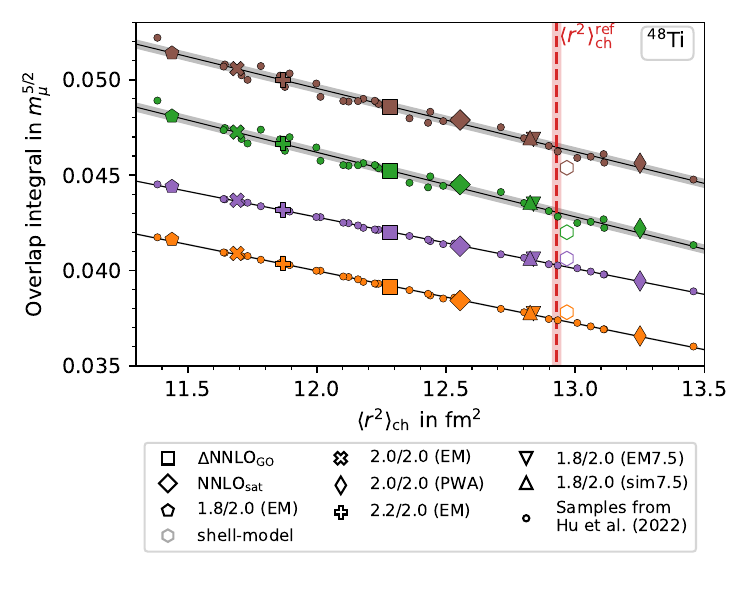}
    \caption{Correlations between $\langle r^2\rangle_\text{ch}$ and the overlap integrals from Eq.~\eqref{eq:S,V,D_kitano} using the IMSRG (specifically VS-IMSRG for $^{27}$Al, $^{48}$Ti) based on a representative set of chiral Hamiltonians (see Appendix~\ref{app:Hamiltonians} for details). The shell-model results, based on Refs.~\cite{Hoferichter:2022mna,Caurier:1999,Caurier:2004gf,Otsuka:2018bqq,Poves:2000nw,Brown:2006gx}, are shown for comparison and are not included in the correlation analysis.}
    \label{fig:Iir2}
\end{figure}%
We compute the structure of nuclei by solving the many-body Schr\"odinger equation using the IMSRG~\cite{Tsukiyama:2010rj, Hergert:2015awm}.
The IMSRG solves for a unitary transformation of the Hamiltonian $U=e^\Omega$ to either directly decouple the ground state from its excitations or alternatively to decouple a core and valence space from the rest of the Hilbert space via the valence-space IMSRG (VS-IMSRG)~\cite{Stroberg:2016ung, Stroberg:2019mxo}, leaving a problem that can be solved using shell-model techniques.
We truncate our (VS-)IMSRG calculations at the level of normal-ordered two-body operators, the (VS-)IMSRG(2),
an approximation that has been demonstrated to be very reliable for ground-state properties of medium-mass nuclei~\cite{Hergert:2015awm, Hergert:2020bxy, Heinz:2024juw}.
All our calculations are performed in an optimized hybrid Hartree--Fock and natural orbital basis following the construction of Ref.~\cite{Heinz:2024juw}, expanded in a basis of 15 major harmonic oscillator shells (with harmonic oscillator frequency $\hbar \omega = 16~\text{MeV}$) before being truncated to an effective model-space size of 11 major shells.
We find the effects of relaxing this truncation to be negligible, indicating that our calculations are converged with respect to model-space size.

Ground-state expectation values of other operators are computed consistently by applying the same unitary transformation.
We compute charge radii based on the point-proton radius operator and the spin-orbit correction~\cite{Hagen:2015yea, Heinz:2024juw},
and we compute the nuclear responses using the $M_{p/n}$ and $\Phi''_{p/n}$ multipole operators from Refs.~\cite{Fitzpatrick:2012ix, Hoferichter:2020osn, Hu:2021awl}.
Both our radius operators and our responses are made translationally invariant through consistent center-of-mass corrections~\cite{Hagen:2009pq, Hagen:2015yea, Sobczyk:2020qtw},
meaning the densities computed as Fourier transforms from our responses give the same $\langle r^2 \rangle$ values as the expectation values of our radius operators. More details on the many-body calculations are provided in Appendix~\ref{app:many_body_uncertainties}.

\section{Overlap integrals}

The overlap integrals in Eq.~\eqref{eq:BR_SI} can be expressed as
\begin{align}
    I_0 &\equiv D = -\frac{4}{\sqrt{2} ~ m_\mu^{3/2}} \int_0^\infty \dd{r} E(r) ~ d(r), \nonumber\\
    I_{1,2} &\equiv S^{(N)} = \frac{\# N}{2\sqrt{2} ~ m_\mu^{5/2}} \int_0^\infty \dd{r} \rho_N(r) ~ s(r), \nonumber\\
    I_{3,4} &\equiv V^{(N)} = \frac{\# N}{2\sqrt{2} ~ m_\mu^{5/2}} \int_0^\infty \dd{r} \rho_N(r) ~ v(r), \label{eq:S,V,D_kitano}
\end{align}
where $\rho_N(r)$ refers to the proton/neutron distribution normalized to 1 with $\#N=Z,A-Z$ for $N=p,n$, respectively, and we put the electron mass $m_e$ to zero (see Ref.~\cite{Noel:2024led} for overlap integrals with finite $m_e$). 
$E(r)$ denotes the electric field of the nucleus, which can be calculated based on the charge distribution $\rho_\text{ch}$, and $s(r)$, $v(r)$, $d(r)$ are combinations of the radial parts of the muon and electron wave functions, which are obtained by numerically solving the Dirac equation with a radial potential given by the electric field. Explicit expressions are given in the Supplementary Material and Refs.~\cite{Noel:2024led, Noel:2024swe}.

Direct calculations of these overlap integrals are challenging, especially for the neutron responses involving $\rho_n$,
as phenomenological approaches employ uncontrolled approximations and current ab initio calculations have considerable uncertainties due to the EFT truncation of the Hamiltonian.
In ab initio calculations, however, these uncertainties are controlled and understood to be strongly correlated in related observables~\cite{Hagen:2015yea, Payne:2019wvy, Heinz:2024juw}, which we leverage by computing the structure of $^{27}$Al, $^{48}$Ca, and $^{48}$Ti for a broad set of chiral EFT Hamiltonians.
We find strong correlations among charge, point-proton, and point-neutron distributions~\cite{Hagen:2015yea,Payne:2019wvy} and exploit these to obtain robust results for the scalar and vector overlap integrals.  As a reference point, we employ the charge densities $\rho^\text{ref}_\text{ch}$ and resulting charge radii squared $\langle r^2\rangle^\text{ref}_\text{ch}$ as extracted from elastic electron scattering in Ref.~\cite{Noel:2024led}.

Figure~\ref{fig:Iir2} shows these correlations between the scalar and vector overlap integrals and the charge radius squared for the three isotopes considered. In each case, a very clear linear relation is observed.
For the proton overlap integrals $S^{(p)}$, $V^{(p)}$, the strong correlation is expected: the point-proton density that enters Eq.~\eqref{eq:S,V,D_kitano} is very closely related to the charge density, so $S^{(p)}$, $V^{(p)}$, and $\langle r^2 \rangle_\text{ch}$ are each computed as weighted integrals involving nearly the same densities.
For this reason, we also find strong correlations between $S^{(p)}$ and $V^{(p)}$ in Appendix~\ref{app:correlation}.
The strong, but slightly more uncertain correlations between $\langle r^2 \rangle_\text{ch}$ and the neutron overlap integrals $S^{(n)}$, $V^{(n)}$ have a more nontrivial origin.
Neutron and proton densities in nuclei are related through nuclear matter saturation properties of nuclear forces~\cite{Hebeler:2015hla, Ekstrom:2015rta}.
Moreover, nuclear forces from chiral EFT themselves are constrained by the symmetries of QCD and the optimization to nucleon-nucleon scattering data~\cite{Epelbaum:2008ga, Machleidt:2011zz},
limiting how much the proton and neutron densities in nuclei can be varied independently (within EFT truncation uncertainties).

We fit each correlation with a simple linear regression according to
\begin{align}
    I_i \qty(\langle r^2 \rangle_\text{ch}) = m_i ~ \qty(\langle r^2 \rangle_\text{ch} - \langle r^2 \rangle^\text{ref}_\text{ch}) + b_i \,. \label{eq:Ii=mi*rsq+bi}
\end{align}
We shift the $x$-axis to be centered around the reference charge radius squared in such a way that ${I_i \qty(\langle r^2 \rangle^\text{ref}_\text{ch}) = b_i}$ and the influence of the correlation between $m_i$ and $b_i$ on the result is minimized. We propagate uncertainties based on the input charge radius squared as well as the fit residuals. We find that the fit residuals, which are due to uncorrelated Hamiltonian and many-body uncertainties in our nuclear structure calculations, appear approximately normally distributed, and for this reason we estimate their uncertainty based on their standard deviation. 
Additionally, we tested our model-space and many-body uncertainties by performing calculations in larger model spaces and using the IMSRG(3)-$N^7$ truncation~\cite{Heinz:2021xir, Heinz:2024juw}, respectively, for a few Hamiltonians in $^{48}$Ca and found that these results also lie perfectly within our established correlations. Based on the extracted charge radii from Ref.~\cite{Noel:2024led}, we find for the overlap integrals the values in Table~\ref{tab:Ii}, reasonably consistent with the results from Ref.~\cite{Kitano:2002mt}, but, crucially, including explicit uncertainty estimates.

\begin{table}[t]%
	\renewcommand{\arraystretch}{1.3}
	\centering
        \scalebox{0.85}{
	\begin{tabular}{l l r r r r r r}
	\toprule
    & $I_i$ & \multicolumn{1}{c}{This work/\cite{Noel:2024led}} & \multicolumn{1}{c}{\cite{Kitano:2002mt}} \\\hline
    \multirow{5}{*}{$^{27}$Al} & $D$ & 0.0359(2) & 0.0362 \\
    & $S^{(p)}$ & 0.01579(2)(19) & 0.0155 \\
    & $S^{(n)}$ & 0.01689(5)(21) & 0.0167 \\
    & $V^{(p)}$ & 0.01635(2)(18) & 0.0161 \\
    & $V^{(n)}$ & 0.01750(5)(21) & 0.0173 \\ \hline
	\multirow{5}{*}{$^{48}$Ca} & $D$ & 0.07479(10) & -- \\
    & $S^{(p)}$ & 0.03265(03)(16) & -- \\
    & $S^{(n)}$ & 0.04250(34)(25) & -- \\
    & $V^{(p)}$ & 0.03483(02)(16) & -- \\
    & $V^{(n)}$ & 0.04561(34)(24) & -- \\ \hline
    \multirow{5}{*}{$^{48}$Ti} & $D$ & 0.08640(11) & 0.0864 \\
    & $S^{(p)}$ & 0.03742(05)(5) & 0.0368 \\
    & $S^{(n)}$ & 0.04305(25)(6) & 0.0435 \\
    & $V^{(p)}$ & 0.04029(04)(5) & 0.0396 \\
    & $V^{(n)}$ & 0.04646(24)(5) & 0.0468 \\
    \bottomrule
	\end{tabular}}
	\caption{Overlap integrals as a result of the correlation analysis in Fig.~\ref{fig:Iir2}. The values for the dipole overlap integral $D$ are taken from Ref.~\cite{Noel:2024led} with the therein quoted total uncertainty. For the other overlap integrals, the first uncertainty component quantifies the remaining nuclear structure uncertainties based on the correlation, see Appendix~\ref{app:correlation}, and the second one is propagated from the reference charge radius squared~\cite{Noel:2024led}.}
	\label{tab:Ii}
\end{table}%

We also determine the correlations among the different overlap integrals with respect to both uncertainty components given in Table~\ref{tab:Ii}. For the uncertainties due to the distributions of the residuals in the correlation, we calculate the pairwise correlation between two residual distributions. For the uncertainties due to the charge radius, we propagate the correlations of the parameters of $\rho_\text{ch}$, which allows us to quantify the correlations with the dipole overlap integrals as well.
Since for $i=1$--$4$ the overlap integrals are all linearly dependent on the same charge radius squared $\langle r^2 \rangle_\text{ch}$, the overlap integrals $I_1$ to $I_4$ are pairwise maximally correlated for the latter uncertainty component. Table \ref{tab:IiIj} shows the combined correlations of both components, clearly displaying the stronger correlations between charge and proton responses and weaker but still sizable correlations between them and the neutron responses. Since the total correlations are a weighted combination of the two uncertainty components, the absolute value of the correlation is strongly dependent on their relative size. If the propagated input uncertainties dominate as in $^{27}$Al, the correlations are significantly stronger than in cases in which the fit uncertainties dominate as in $^{48}$Ti, since those are statistically more independent.

\begin{table}[t]
\centering
	\renewcommand{\arraystretch}{1.3}
    \scalebox{0.85}{
    \begin{tabular}{l r r r r r}
	\toprule
    \multicolumn{6}{c}{$^{27}$Al} \\\hline
     & $D$ & $S^{(p)}$ & $S^{(n)}$ & $V^{(p)}$ & $V^{(n)}$\\\hline
    $D$ & 1.0000 & 0.7205 & 0.7030 & 0.7210 & 0.7028\\
    $S^{(p)}$ &  & 1.0000 & 0.9656 & 1.0000 & 0.9645\\
    $S^{(n)}$ &  &  & 1.0000 & 0.9664 & 1.0000\\
    $V^{(p)}$ &  &  &  & 1.0000 & 0.9654\\
    $V^{(n)}$ &  &  &  &  & 1.0000\\
	\bottomrule
	\end{tabular}}
    \vspace{2mm}
    
    \scalebox{0.85}{
    \begin{tabular}{l r r r r r}
	\toprule
    \multicolumn{6}{c}{$^{48}$Ca} \\\hline
     & $D$ & $S^{(p)}$ & $S^{(n)}$ & $V^{(p)}$ & $V^{(n)}$\\\hline
    $D$ & 1.0000 & 0.8938 & 0.5295 & 0.8956 & 0.5272\\
    $S^{(p)}$ &  & 1.0000 & 0.6125 & 0.9999 & 0.6089\\
    $S^{(n)}$ &  &  & 1.0000 & 0.6120 & 0.9999\\
    $V^{(p)}$ &  &  &  & 1.0000 & 0.6085\\
    $V^{(n)}$ &  &  &  &  & 1.0000\\
	\bottomrule
	\end{tabular}}
    \vspace{2mm}
    
    \scalebox{0.85}{
    \begin{tabular}{l r r r r r}
	\toprule
    \multicolumn{6}{c}{$^{48}$Ti} \\\hline
     & $D$ & $S^{(p)}$ & $S^{(n)}$ & $V^{(p)}$ & $V^{(n)}$\\\hline
    $D$ & 1.0000 & 0.4657 & 0.1169 & 0.5003 & 0.1163\\
    $S^{(p)}$ &  & 1.0000 & 0.1118 & 0.9991 & 0.0916\\
    $S^{(n)}$ &  &  & 1.0000 & 0.1176 & 0.9997\\
    $V^{(p)}$ &  &  &  & 1.0000 & 0.0978\\
    $V^{(n)}$ &  &  &  &  & 1.0000\\
	\bottomrule
	\end{tabular}}
    
	\caption{Total correlations among the different overlap integrals as a combination of the propagated correlations from the observed correlations among the fit residuals and the reference charge density. The number of digits quoted does not represent precision but is chosen for reproducibility such that the eigenvalues of the correlation matrix remain non-negative.}
	\label{tab:IiIj}
\end{table}

\section{Neutron skin}

Our work allows us to predict weak scattering in nuclei in a controlled and precise way. To test our predictions against existing weak-scattering data, we exploit the same correlations among the charge radius squared and the point-proton, point-neutron, and weak radius squared to extract values for these radii based on the input radius $\langle r^2 \rangle^\text{ref}_\text{ch}$ from Ref.~\cite{Noel:2024led}. We compare the resulting neutron and weak skin thickness to the results from the PVES experiments $Q_\text{weak}$ on $^{27}$Al~ \cite{Qweak:2021ijt} and CREX on $^{48}$Ca~\cite{CREX:2022kgg} as shown in Table~\ref{tab:r&F}. We find mostly consistent values with some very slight tensions, which mostly trace back to different input values for the charge radius. We also compare to the correlation analysis of Ref.~\cite{Hagen:2015yea}, which shows good consistency despite a much smaller set of chiral interactions and the use of a large proton radius in the conversion between charge and point-proton radii~\cite{Hoferichter:2020osn}. We emphasize that through the use of a large ensemble of Hamiltonians and a detailed correlation analysis including also many-body uncertainties we improve upon past work and are able make substantially more precise predictions for weak-scattering properties. This can further refine the analysis and interpretation of PVES experiments. 

PVES experiments measure the left--right asymmetry~\cite{PREX:2021umo, Qweak:2021ijt, CREX:2022kgg}, which becomes proportional to the weak form factor at the respective momentum transfer only in the plane-wave limit, while a rigorous extraction requires the consideration of Coulomb corrections. As the weak density is not fully known, this inevitably introduces some model dependence. For this reason, we also consider the direct correlation between charge and weak form factor at the momentum transfer of the experiment, using as reference value the form factor calculated via the charge distribution from Ref.~\cite{Noel:2024led}. The resulting values are also listed in Table~\ref{tab:r&F}, together with the respective results from $Q_\text{weak}$ and CREX. 

For $^{27}$Al our result is roughly 8\,\% smaller than the experimental one, which is amply covered by the $\simeq 10\,\%$ uncertainty on the experimental value and can be attributed, to a large part,  to the smaller charge form factor used as input. For $^{48}$Ca the difference is a bit larger and the experimental uncertainty is significantly smaller, such that at face value we see a tension around $2\sigma$. However, part of the tension again originates from different input for the charge form factor.  Further, if we calculate the left--right asymmetry (including Coulomb corrections) based on our reference charge density and the extracted weak density (see Supplementary Material for details), we find an asymmetry that comes significantly closer to the experimental result, reducing the mismatch to about $1\sigma$. This observation suggests that the result for the weak form factor as extracted by CREX might depend more strongly on the details of the calculation of the Coulomb corrections than assumed in Ref.~\cite{CREX:2022kgg}.
\begin{table}[t]%
	\renewcommand{\arraystretch}{1.3} 
	\centering  
    \scalebox{0.85}{
    \begin{tabular}{l l r r r r r r} 
	\toprule 
    & & & \multicolumn{1}{c}{This work/\cite{Noel:2024led}} & References \\ \hline
    \multirow{4}{*}{$^{27}$Al} 
    & $r_\text{n}-r_\text{p}$ & $[\text{fm}]$ & $0.021(09)(46)$ & $-0.04(12)$ & \cite{Qweak:2021ijt} \\
    & $r_\text{w}-r_\text{ch}$ & $[\text{fm}]$ & $0.023(10)(44)$ & $-0.04(15)$ & \cite{Qweak:2021ijt} \\
    & $F^\text{exp}_\text{ch}$ & & $0.3665(45)$ & $0.382(12)$ & \cite{Qweak:2021ijt} \\
    & $F^\text{exp}_\text{w}$ & & $0.3614(23)(46)$ & $0.393(38)$ & \cite{Qweak:2021ijt} \\ \hline
    \multirow{5}{*}{$^{48}$Ca} 
    & \multirow{2}{*}{$r_\text{n}-r_\text{p}$} & \multirow{2}{*}{$[\text{fm}]$} & \multirow{2}{*}{$0.152(17)(15)$} & [0.12,\,0.15] & \cite{Hagen:2015yea} \\ &  & &  & $0.121(26)(24)$ & \cite{CREX:2022kgg} \\
    & $r_\text{w}-r_\text{ch}$ & $[\text{fm}]$ & $0.191(18)(15)$ & $0.159(26)(23)$ & \cite{CREX:2022kgg} \\
    & $F^\text{exp}_\text{ch}$ & & $0.15603(55)$ & $0.1581$ & \cite{CREX:2022kgg} \\
    & $F^\text{exp}_\text{w}$ & & $0.1171(28)(5)$ & $0.1304(52)(20)$ & \cite{CREX:2022kgg} \\ 
    \bottomrule 
	\end{tabular}} 
    
	\caption{Neutron and weak skin calculated as the difference between the respective radii that were correlated to the charge radius (for individual radii and the results for $^{48}$Ti see Supplementary Material). We also considered correlating the radius differences directly, leading to identical central values, but a reduced sensitivity to the reference charge radius. 
    We further provide the charge and weak form factor at the momentum transfer used by the respective PVES experiments $F^\text{exp}_\text{ch,w} = F_\text{ch,w}(q_\text{exp})$ with $q_\text{exp}=0.87335(58)\,\text{fm}^{-1}$\cite{CREX:2022kgg} for $^{48}$Ca and $q_\text{exp}=0.77802(33)\,\text{fm}^{-1}$\cite{Qweak:2021ijt} for $^{27}$Al. The uncertainties propagated from the spread in the momentum are not listed in the table, but amount to roughly $0.0005$ for CREX and $0.0004$ for $Q_\text{weak}$ in the quoted weak form factors. In all cases the first uncertainty component quantifies the remaining nuclear structure uncertainties based on the correlation, see Appendix~\ref{app:correlation}, and the second one is propagated from the reference charge radius squared~\cite{Noel:2024led}.}
	\label{tab:r&F}

\end{table}

\section{Conclusions}

In this work, we calculated the overlap integrals for $\mu\to e$ conversion in nuclei corresponding to the leading SI responses using ab initio methods, providing, for the first time, robust uncertainty estimates including correlations among the different integrals. To this end, we explored correlations among the overlap integrals and the nuclear charge radius for the phenomenologically most relevant isotopes $^{27}$Al and $^{48}$Ti as well as $^{48}$Ca, the latter allowing for validation against previous work and data from PVES. As our main result, see Fig.~\ref{fig:Iir2}, we observe that tight correlations exist even for the neutron responses, covering both a variety of chiral Hamiltonians and many-body uncertainties.
We exploit these correlations to control for common systematic uncertainties in our calculations, providing much more stringent constraints on neutron densities and related observables than in previous work.
This is to be contrasted with a probabilistic treatment of systematic uncertainties~\cite{Furnstahl:2015rha,Melendez:2019izc,Phillips:2020dmw,Hu:2021trw}, which would require conditional error modeling to properly account for the correlated nature of these uncertainties.
We are additionally able to model the remaining uncertainties in the correlation analysis in an approximate statistical way.
This makes it now possible to propagate nuclear uncertainties in the evaluation of the $\mu\to e$ conversion rate in Eq.~\eqref{eq:BR_SI}, crucial for any robust assessment of the sensitivity to different underlying LFV mechanisms. Similar strategies will allow for improved calculations of nuclear matrix elements for PVES, coherent neutrino--nucleus scattering~\cite{Abdullah:2022zue,Ruso:2022qes}, and the direct detection of dark matter~\cite{Aalbers:2022dzr,XLZD:2024nsu}.

\section*{Acknowledgments} 
We thank Baishan Hu for valuable discussions.
Financial support by the Swiss National Science Foundation (Project No.\ TMCG-2\_213690) is gratefully acknowledged.
This work was supported in part by the European Research Council (ERC) under the European Union's Horizon 2020 research and innovation programme (Grant Agreement No.~101020842),
by the U.S.\ Department of Energy, Office of Science, Office of Advanced Scientific Computing Research and Office of Nuclear Physics, Scientific Discovery through Advanced Computing (SciDAC) program (SciDAC-5 NUCLEI), 
by the Laboratory Directed Research and Development Program of Oak Ridge National Laboratory, managed by UT-Battelle, LLC, for the U.S.\ Department of Energy,
and by JST ERATO Grant No.~JPMJER2304, Japan.
This research used resources of the Oak Ridge Leadership Computing Facility located at Oak Ridge National Laboratory, which is supported by the Office of Science of the Department of Energy under contract No.~DE-AC05-00OR22725.
The authors gratefully acknowledge the Gauss Centre for Supercomputing e.V.\ (www.gauss-centre.eu) for funding this project by providing computing time through the John von Neumann Institute for Computing (NIC) on the GCS Supercomputer JUWELS at J\"ulich Supercomputing Centre (JSC).

This manuscript has been authored in part by UT-Battelle, LLC, under contract DE-AC05-00OR22725 with the US Department of Energy (DOE). The US government retains and the publisher, by accepting the article for publication, acknowledges that the US government retains a nonexclusive, paid-up, irrevocable, worldwide license to publish or reproduce the published form of this manuscript, or allow others to do so, for US government purposes. DOE will provide public access to these results of federally sponsored research in accordance with the DOE Public Access Plan (\url{http://energy.gov/downloads/doe-public-access-plan}).

\appendix

\section{Ensemble of nuclear Hamiltonians}
\label{app:Hamiltonians}

We use an ensemble of Hamiltonians with NN and 3N potentials from chiral EFT in our (VS-)IMSRG calculations to explore EFT truncation uncertainties.
The Hamiltonians we employ vary in their truncation order, their regularization cutoff scales, details of their EFT construction, and how they are fit to data.
All Hamiltonians are fit to NN scattering data, deuteron properties, and properties of few-body systems with $A\leq4$, but some are optimized against additional constraints. The long-range parts are constrained using low-energy constants (LECs) determined from pion--nucleon scattering~\cite{Hoferichter:2015tha,Hoferichter:2015hva,Siemens:2016jwj}.
Furthermore, some Hamiltonians are transformed to lower resolution scales using the similarity renormalization group (SRG)~\cite{Bogner:2006pc}, making them more perturbative and amenable to many-body calculations.

\begin{table}[t]
\renewcommand{\arraystretch}{1.3}
    \centering
    \begin{threeparttable}
    \scalebox{0.85}{
    \begin{tabular}{lccccc}
    \toprule
    \multirow{2}{*}{Name} & \multirow{2}{*}{Ref.} & \multicolumn{2}{c}{Order, Cutoff} & \multirow{2}{*}{$\lambda$} & Optimized \\
    & & NN & 3N & & to $A > 4$? \\
    \hline
    \hebelerA{} & \multirow{4}{*}{\cite{Hebeler:2010xb}} & \multirow{4}{*}{\nnnlo{}, 500} & \multirow{4}{*}{\nnlo{}, 394} & 1.8 & \multirow{4}{*}{No} \\
    \hebelerB{} & & & & 2.0 &  \\
    \hebelerC{} & & & & 2.2 &  \\
    \hebelerD{} & & & & 2.0 &  \\
    \hline
    \arthuisA{} & \multirow{2}{*}{\cite{Arthuis:2024mnl}} & \nnnlo{}, 500 & \multirow{2}{*}{\nnlo{}, 394} & 1.8 & \multirow{2}{*}{Yes} \\
    \arthuisB{} & & \nnlo{}, 550 &  & 1.8 & \\ 
    \hline
    \nnlosat{} & \cite{Ekstrom:2015rta} & \nnlo{}, 450 & \nnlo{}, 450 & -- & Yes \\
    \hline
    \dnnlogo{}\tnote{2} & \cite{Jiang:2020the} & \nnlo{}, 394 & \nnlo{}, 394 & -- & \hphantom{\footnotemark[1]}Yes\tnote{1} \\
    \hline
    34 samples & \multirow{2}{*}{\cite{Hu:2021trw}} & \multirow{2}{*}{\nnlo{}, 394} & \multirow{2}{*}{\nnlo{}, 394} & \multirow{2}{*}{--} & \multirow{2}{*}{Yes} \\ 
    from Hu et al.\tnote{2}& \\
    \bottomrule
    \end{tabular}}
    {\footnotesize
    \begin{tablenotes}
  \item[1] Fit to nuclear matter properties.
  \item[2] Explicit inclusion of $\Delta$ isobars in EFT construction.
  \end{tablenotes}
  }
    \caption{
        \label{tab:hams_table}
        Nuclear Hamiltonians from chiral EFT used in this work and in Fig.~\ref{fig:Iir2}.
        Regulator cutoffs are given in MeV.
        SRG resolution scales $\lambda$ are given in \fmi{} where relevant.
        Hamiltonians optimized to $A > 4$ are generally fit to ground-state energies and charge radii of $^{16}$O, but occasionally also selected ground-state energies and charge radii of other nuclei such as $^{14}$C and $^{22,24,25}$O.
    }
    \end{threeparttable}
\end{table}

An overview of all 42 nuclear Hamiltonians we use is given in Table~\ref{tab:hams_table}~\cite{Hebeler:2010xb, Ekstrom:2015rta, Jiang:2020the, Hu:2021trw, Arthuis:2024mnl}.
They are generally at next-to-next-to-leading order (\nnlo), but selected Hamiltonians have NN interactions at one order higher, \nnnlo{}.
The Hamiltonians from Refs.~\cite{Jiang:2020the,Hu:2021trw} explicitly include $\Delta$ isobars in their chiral EFT construction.
Regulator cutoffs range from 394 to 550~\MeV{},
and low-resolution Hamiltonians with SRG resolution scales from $\lambda = 1.8$ to 2.2~\fmi{} are explored.
The 34 interactions of Ref.~\cite{Hu:2021trw} are samples from distributions of LECs determined through a history matching procedure comparing against NN scattering data, deuteron properties, ground-state energies and charge radii of few-body systems with $A\leq 4$, and the ground-state energy and charge radius of $^{16}$O, capturing the Hamiltonian uncertainty through uncertainties in the underlying LECs.\footnote{
For our calculations of $^{27}$Al, we find that six of the 34 samples from Ref.~\cite{Hu:2021trw} give unphysical results in our calculations. These are not shown and excluded from our analysis.
}
The broad range of Hamiltonians we consider allows us to probe many aspects of the chiral EFT truncation uncertainty, and the strong correlations we find apply to all interactions, indicating that exploiting such correlations is insensitive to specific details of the Hamiltonian construction.

\begin{figure}[t]%
    \centering
    \includegraphics[width=0.9\linewidth,trim=0 10 0 10,clip]{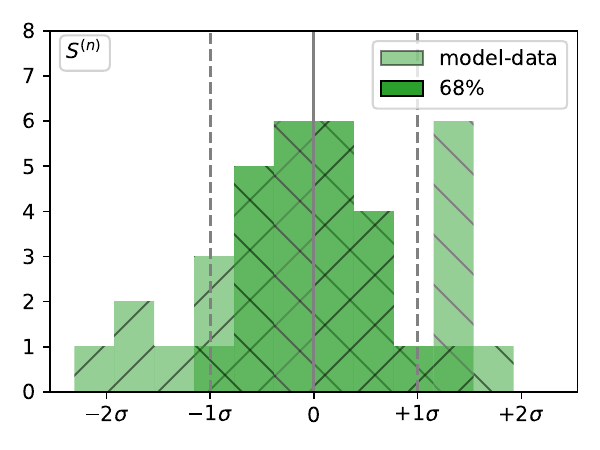}
    \caption{Residual distribution for the $S^{(n)}$ overlap integral of $^{27}$Al.
    The dark (light) region marks the residual values that are in (outside) 68\,\% of the closest residuals to zero.
    }
    \label{fig:Al27_resid_dist}
\end{figure}

\begin{figure}[t]%
    \centering
    \includegraphics[width=0.99\linewidth, trim = 0 10 35 40,clip]{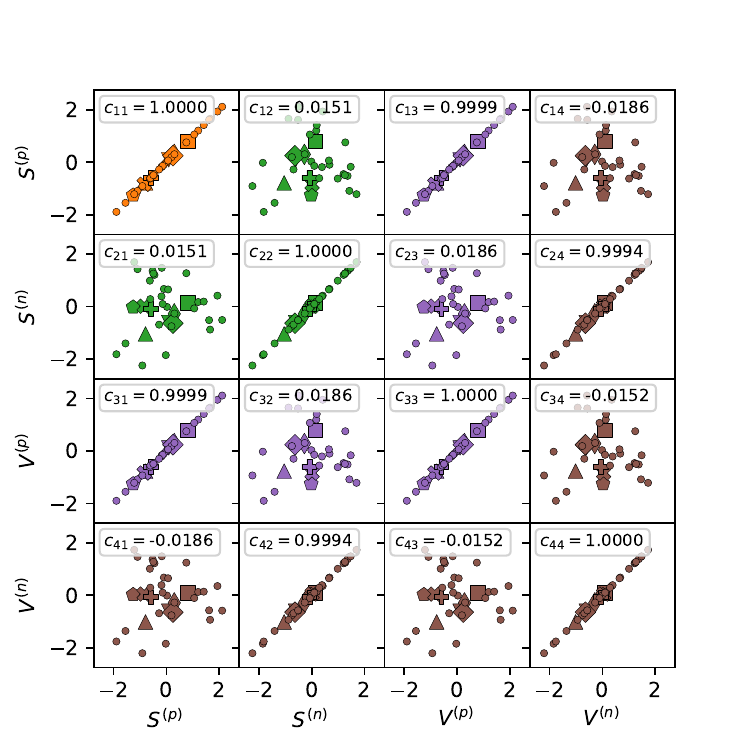}
    \caption{Correlations of the residuals for $^{27}$Al normalized to the uncertainty based on the distribution as shown in Fig.~\ref{fig:Al27_resid_dist}, using the same symbols as in Fig.~\ref{fig:Iir2}. $c_{ij}$ refers to the correlation between the residuals of $I_i$ and $I_j$.}
    \label{fig:Al27_resid_corr}
\end{figure}

\section{Many-body uncertainties}
\label{app:many_body_uncertainties}

In addition to the EFT uncertainties explored by our ensemble of Hamiltonians, ab initio many-body calculations also have model-space uncertainties due to working in a finite basis and many-body uncertainties due to truncations in the many-body method.
For the \hebelerA{}, \nnlosat{}, and \arthuisA{} Hamiltonians, we explored these uncertainties in $^{48}$Ca by increasing our model-space size to 13 effective harmonic oscillator shells
and by performing IMSRG(3)-$N^7$ calculations with restricted normal-ordered three-body operators~\cite{Heinz:2024juw}, improving on the IMSRG(2) truncation we employ for the rest of this work.
For both the larger model-space size and the more precise IMSRG(3)-$N^7$ truncation, the resulting charge radii and overlap integrals in Fig.~\ref{fig:Iir2} shift by small amounts essentially exactly along the linear correlations we find.
This indicates that many-body uncertainties are similarly correlated in ab initio calculations of point-proton, point-neutron, and charge densities.
Specifically, the correlations we find can be applied without considering additional uncorrelated uncertainties due to truncations and approximations in our (VS-)IMSRG calculations.

\section{Details of correlation analysis}
\label{app:correlation}

The correlation analysis employs a linear fit according to Eq.~\eqref{eq:Ii=mi*rsq+bi} using the Levenberg--Marquardt method~\cite{Levenberg:1944,Marquardt:1963} as implemented by the \texttt{python} package \texttt{lmfit}~\cite{lmfit:2024}. We consider two uncertainty components: one propagated from the input quantities and one based on the fit quality. The former is straightforward to implement, using the uncertainty of $\langle r^2 \rangle^\text{ref}_\text{ch}$ coming from the parameterization of $\rho^\text{ref}_\text{ch}$, and propagating it according to Eq.~\eqref{eq:Ii=mi*rsq+bi}.\footnote{We follow the preferred strategy for the estimate of systematic uncertainties as argued in Ref.~\cite{Noel:2024led}.
This implies that for $^{48}$Ti the result is more strongly constrained by the charge radius extracted from muon spectroscopy, in contrast to $^{27}$Al and $^{48}$Ca, in which cases the uncertainties assigned in Fig.~\ref{fig:Iir2} account for the tension observed between scattering data and spectroscopy as well.}  For the latter, an uncertainty estimate solely based on the fit statistics will inevitably grossly underestimate the resulting uncertainties, particularly if the uncertainties of the individual data points are unknown and sizable correlations among different results from the same chiral EFT scheme cannot be included systematically. 
We observe, however, that the residuals of the fit appear approximately normally distributed. For this reason we define the uncertainty in a purely statistical sense based on the residual distribution of the $b_i$ and thus the overlap integrals based on the distribution of the residuals in such a way that $\Delta I_i = \Delta b_i$ is given by the standard deviation of all residuals of the fit. This approach provides a simple way to estimate the correlation uncertainties, including all the aforementioned systematic effects implicitly, and propagate them to other results. Figure~\ref{fig:Al27_resid_dist} shows exemplarily such a residual distribution for the $S^{(n)}$ overlap integral for $^{27}$Al and confirms that the resulting $1\sigma$ uncertainties are meaningful, as they coincide, within reason, with the distance of 68\,\% of the residuals.
The correlations of this uncertainty component among the different overlap integrals can be calculated with statistical methods based on the different residual distributions. We find the correlations as illustrated in Fig.~\ref{fig:Al27_resid_corr} for $^{27}$Al, which shows the observed strong correlation between overlap integrals that couple to the same nuclear density.
This correlation is then combined with the propagated correlation from the input charge density parametrization, resulting in the values from Table~\ref{tab:IiIj}. We find similar behavior for the other nuclei studied in this work (see Supplementary Material~\cite{SuppMatt} with Refs.~\cite{Romao:2012pq, Hoferichter:2018acd, Hoferichter:2016nvd, DeVries:1987atn, Angeli:2013epw}).

\bibliographystyle{apsrev4-1_mod}
\balance
\biboptions{sort&compress}
\bibliography{ref}

\clearpage

\onecolumn

\setcounter{figure}{3}
\setcounter{table}{4}
\setcounter{section}{0}

\begin{center}
{\Large Supplementary Material to\\``Ab initio calculations of overlap integrals for $\mu\to e$ conversion in nuclei''}

\end{center}

\begin{multicols}{2}

\section*{\texorpdfstring{$\mu\to e$}{} conversion} \label{app:mutoe}

The leading coherently enhanced contributions to $\mu\to e$ conversion come from scalar, vector, and dipole interactions. The relevant effective operators up to dimension seven are given by
\begin{align}
	\mathcal{L}^\text{SI}_{\text{eff}} &= \frac{1}{\Lambda^2} \sum_{\substack{Y=L,R\\q=u,d,s}}\! \Big[ C^{S,q}_Y (\overline{e_Y} \mu)(\bar{q} q) + C^{V,q}_Y (\overline{e_Y} \gamma^\mu \mu)(\bar{q} \gamma_\mu q) \Big] \nonumber\\
    & \quad + \frac{1}{\Lambda} \sum_{\substack{Y=L,R}} C^D_Y (\overline{e_Y} \sigma^{\mu\nu} \mu) F_{\mu\nu} \nonumber\\
    & \quad + \frac{\alpha_s}{\Lambda^3} \sum_{\substack{Y=L,R}}\! C^{GG}_Y (\overline{e_Y} \mu) (G^a_{\alpha\beta} G_a^{\alpha\beta}) ~ + ~ \text{h.c.}~, \label{eq:QuarkLevelLagrangian}
\end{align}
where we use $\overline{e_Y}=\overline{e P_Y}=\overline{e}P_{\bar{Y}}$ with $Y \in \{L,R\}$ and $P_{L/R} = (\Id \mp \gamma_5)/2$. This convention makes the decoupling of the left-handed and right-handed components of the electron for $m_e\to0$ explicit. We introduce the BSM scale $\Lambda$ to make the Wilson coefficients $C^{X(,q)}_Y$ with ${X=S,V,D,GG}$ dimensionless. Using these conventions, the SI $\mu\to e$ conversion rate takes the form quoted in Eq.~(2) of the main text with the prefactors given as~\cite{Hoferichter:2022mna} 
\begin{align}
    \bar{C}^{S^{(N)}}_Y &= \frac{1}{\Lambda^2}\sum_q C^{S,q}_{Y} \frac{m_N}{m_q} f^{N}_q + \frac{4 \pi}{\Lambda^3} C^{GG}_Y a_N,\notag\\
    \bar{C}^{V^{(N)}}_Y &= \frac{1}{\Lambda^2}\sum_q C^{V,q}_{Y} f^N_{V_q}, \notag\\
    \bar{C}^{D}_Y &= \frac{\eta_e}{4 m_\mu \Lambda} C^{D}_{Y},
    \label{coefficients_SI}
\end{align}
where $\eta_e$ refers to the sign convention for the charge, corresponding to a  minimal coupling of ${D_\mu=\partial_\mu + i \eta_e \sqrt{4\pi\alpha_\text{el}} A^\mu}$~\cite{Romao:2012pq}, ${\alpha_\text{el}=e^2/(4\pi)}$,
and the hadronic matrix elements defined by
\begin{align}
	\bra{N} m_q \bar{q} q \ket{N} &= \bar{u}_{N}' \big[m_N f^N_q(q)\big] u_{N}, \nonumber\\
	\bra{N} \bar{q} \gamma^\mu q \ket{N} &= \bar{u}_{N}' \Big[\gamma^\mu F^{q,N}_1(q) - \frac{i \sigma^{\mu\nu}q_\nu}{2 m_N} F_2^{q,N}(q)\Big] u_{N}, \nonumber\\
    \bra{N} G^a_{\mu\nu} G_a^{\mu\nu} \ket{N} &= \bar{u}_{N}' \Big[\frac{4\pi}{\alpha_s} a_N(q)\Big] u_{N}, 
	\label{eq:matrixelements_nucleons}
\end{align}
with $\bar{u}_{N}=\bar{u}_{N}(p,s)$, $\bar{u}_{N}'=\bar{u}_{N}(p',s')$, momenta and spins of initial and final nucleon $p,s$ and $p',s'$, and $q=p-p'$, see, e.g., Refs.~\cite{Hoferichter:2022mna,Hoferichter:2020osn,Hoferichter:2018acd,Hoferichter:2016nvd} for a review of the required matrix elements. 

The overlap integrals are defined in Eq.~(3) of the main text, with the  electric field of the nucleus given by 
\begin{align}
    E(r) &= \frac{\sqrt{4\pi \alpha_\text{el}}}{r^2} \int_0^r \dd{r'} r'^2 \rho_\text{ch}(r'),
\end{align}
in terms of the charge distribution $\rho_\text{ch}$. The combinations of the radial parts of the muon and electron wave functions are defined as 
\begin{align}
    s(r) &= g^{e}_{-1}(r) \, g^{\mu}_{-1}(r) - f^{e}_{-1}(r) \, f^{\mu}_{-1}(r), \nonumber\\ 
    v(r) &= g^{e}_{-1}(r) \, g^{\mu}_{-1}(r) + f^{e}_{-1}(r) \, f^{\mu}_{-1}(r), \nonumber\\
    d(r) &= g^{e}_{-1}(r) \, f^{\mu}_{-1}(r) + f^{e}_{-1}(r) \, g^{\mu}_{-1}(r), \label{eq:s,v,d}
\end{align}
where the full wave functions are decomposed as
\begin{align}
    \psi_\kappa^\mu(\rr)= \frac{1}{r} \mqty( g_\kappa(r) \phi_\kappa^\mu(\rrhat) \\ i f_\kappa(r) \phi_{-\kappa}^\mu(\rrhat)), \label{eq:num_wavefct}
\end{align}
separating the angular-momentum degrees of freedom into $\phi_\kappa^\mu(\rrhat)$. The quantum numbers are contained in $\kappa\gtrless 0$ according to
\begin{align}
    j &= \abs{\kappa} - \frac{1}{2} ,& j_z &= \mu, & 
    l &= \begin{cases} \kappa &  \quad \kappa>0, \\ -\kappa-1 &  \quad \kappa<0. \end{cases} \label{eq:j,l,kappa}
\end{align}
The radial potential of the nucleus necessary for solving the Dirac equation is given by the electric field as
\begin{align}
    V(r)&=-\sqrt{4\pi \alpha_\text{el}} \int_r^\infty \dd{r'} E(r').
\end{align}
Again, we refer to Refs.~\cite{Noel:2024led, Noel:2024swe} for further details and conventions.

\end{multicols}
\twocolumn

\section*{Proton, neutron, and weak distributions}
\label{app:distributions}

\begin{figure}
    \centering
    \includegraphics[width=0.99\linewidth, trim = 20 0 20 0]{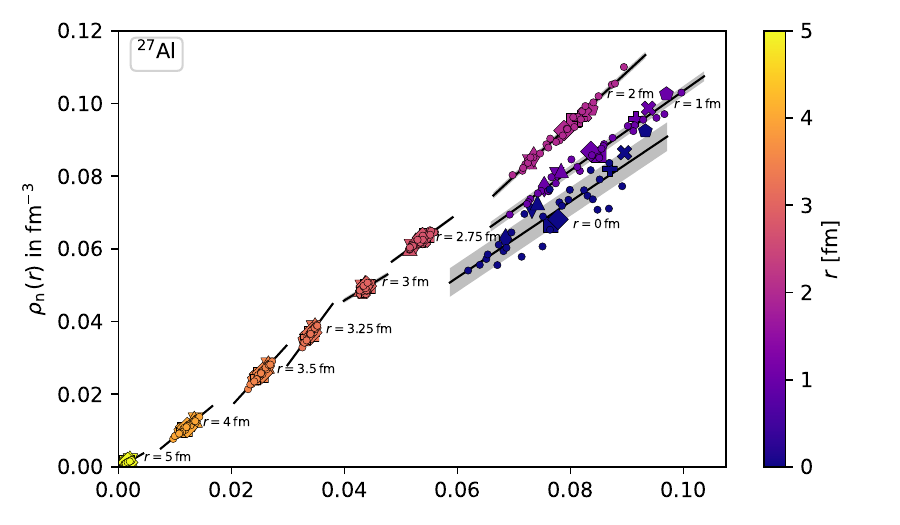}\\[-0.3cm]
    \includegraphics[width=0.99\linewidth, trim = 20 0 20 0]{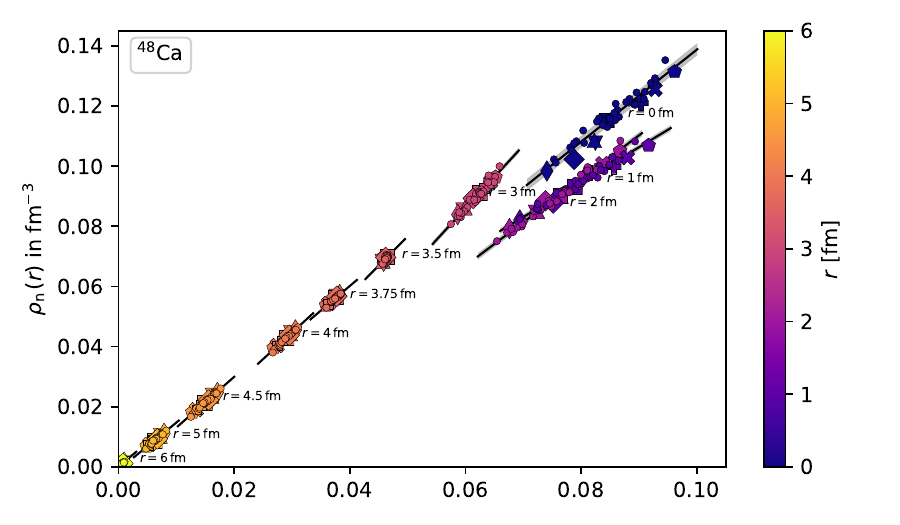}\\[-0.3cm]
    \includegraphics[width=0.99\linewidth, trim = 20 0 20 0]{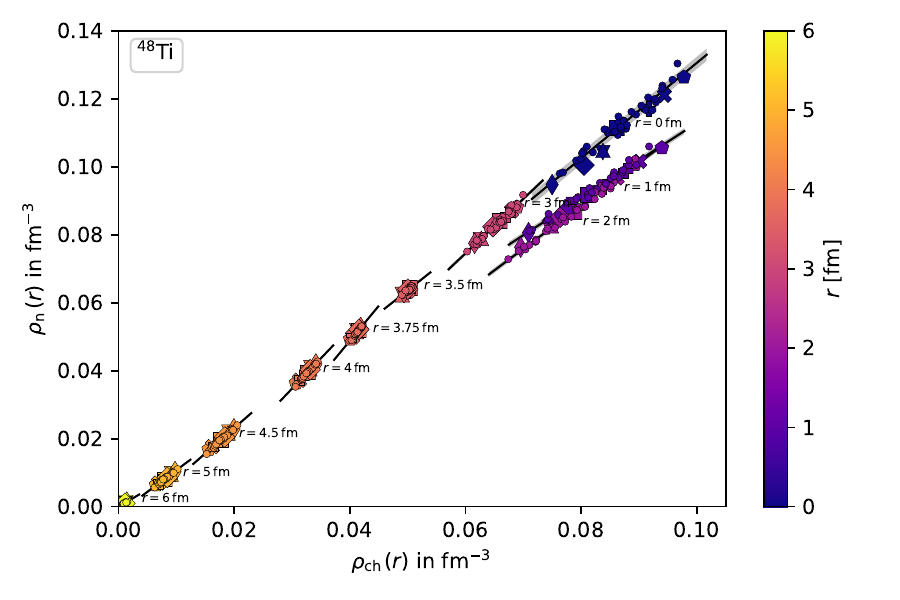}\\[-0.3cm]
    \caption{Correlations between $\rho_\text{ch}$ and $\rho_\text{n}$ illustrated for selected values of $r$ for $^{27}$Al, $^{48}$Ca, and $^{48}$Ti. Symbols are as used in Fig.~1 of the main text.}
    \label{fig:corr_rho}
\end{figure}

\begin{figure}
    \centering
    \includegraphics[width=0.99\linewidth,trim= 0 5 0 0]{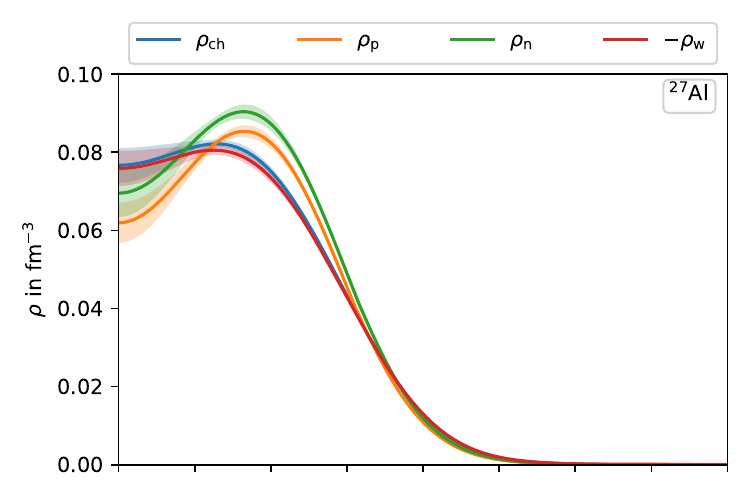} 
    \includegraphics[width=0.99\linewidth,trim= 0 5 0 5]{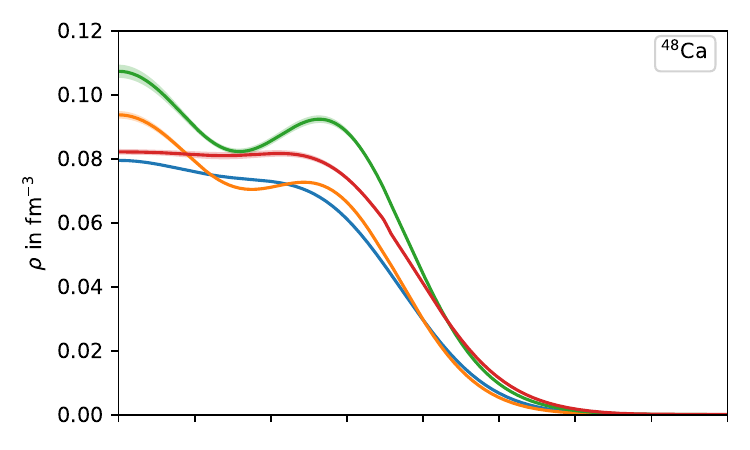} 
    \includegraphics[width=0.99\linewidth,trim= 0 5 0 5]{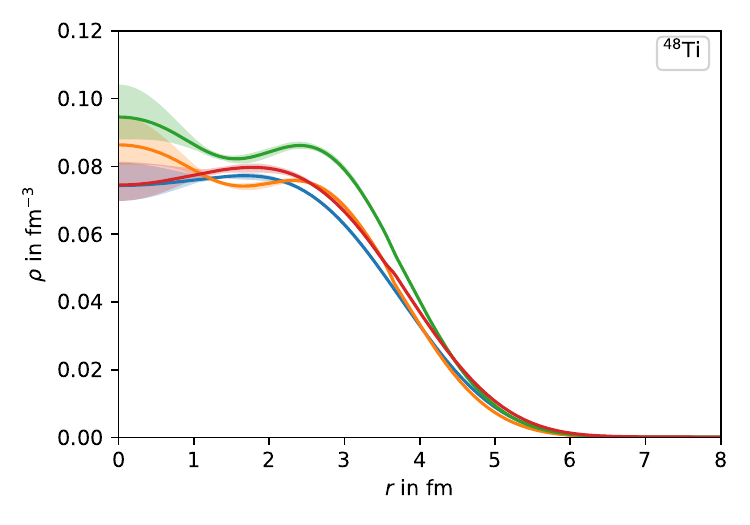} 
    \caption{Point-proton, point-neutron, and weak distributions for $^{27}$Al, $^{48}$Ca, and $^{48}$Ti based on a correlation analysis at fixed $r$ together with the input charge density; $\rho_\text{ch}=\rho^\text{ref}_\text{ch}$.}
    \label{fig:densities}
\end{figure}

We can also perform the same correlation exercise individually between $\rho_\text{ch}(r)$ and $\rho_\text{p}(r)$, $\rho_\text{n}(r)$, or $\rho_\text{w}(r)$ directly. While this seems to work quite well empirically, we note that it is not clear that fitting the correlations at different $r$ completely independently does not introduce additional systematic effects. Figure~\ref{fig:corr_rho} shows the linear trend for a selection of values for $r$, scanning over the different interactions, for the correlation between $\rho_\text{ch}(r)$ and $\rho_\text{n}(r)$. With a fine sampling of $r$ we can calculate distributions $\rho_\text{p}(r)$, $\rho_\text{n}(r)$, and $\rho_\text{w}(r)$ based on the extracted correlations and $\rho^\text{ref}_\text{ch}(r)$. We find the results shown in Fig.~\ref{fig:densities}, where we included the uncertainties in the same way as for the overlap integrals. The uncertainties increase towards smaller radii, where the nucleus is probed less precisely experimentally. Moreover, as expected, the uncertainties for the weak and neutron distributions are larger than for the proton ones due to the higher uncertainties coming from the spread of the considered interactions. As a crosscheck, we calculate the radii based on these distributions. We find, with a rough uncertainty estimate just based on the upper and lower error bands,
\begin{align}
    r^{^{27}\text{Al}}_\text{p} &= 3.0(2)\,\text{fm}, &  r^{^{27}\text{Al}}_\text{n} &= 3.0(2)\,\text{fm}, \notag\\
    r^{^{27}\text{Al}}_\text{w} &= 3.0(2)\,\text{fm}, & &\notag\\
    r^{^{48}\text{Ca}}_\text{p} &= 3.40(4)\,\text{fm}, &  r^{^{48}\text{Ca}}_\text{n} &= 3.55(8)\,\text{fm},\notag\\
     r^{^{48}\text{Ca}}_\text{w} &= 3.67(10)\,\text{fm}, && \notag\\
    r^{^{48}\text{Ti}}_\text{p} &= 3.49(2)\,\text{fm}, &  r^{^{48}\text{Ti}}_\text{n} &= 3.55(3)\,\text{fm},\notag\\
    r^{^{48}\text{Ti}}_\text{w} &= 2.68(3)\,\text{fm}, & &
\end{align}
which are all consistent with the results from Table~\ref{tab:r2}, but give significantly larger uncertainties due to the simple methodology of the uncertainty estimate here for which, in particular, the total charge was not constrained. Nevertheless, these values serve as a valuable crosscheck that the extracted distributions are meaningful.

\clearpage

\begin{figure}[p]
    \centering
    \includegraphics[width=0.95\linewidth,trim=0 10 0 10,clip]{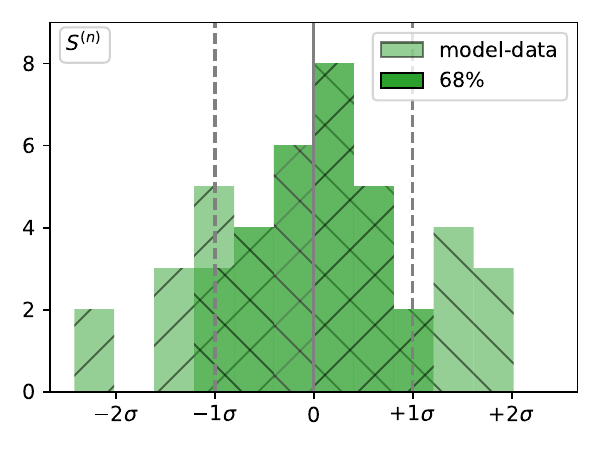}
    \caption{Residual distribution for the $S^{(n)}$ overlap integral of $^{48}$Ca. The dark (light) region marks the residual values that are in (outside) 68\,\% of the closest residuals to zero.}
    \label{fig:Ca48_resid_dist}
\end{figure}%

\begin{figure}[p]
    \centering
    \includegraphics[width=0.95\linewidth,trim=0 10 0 10,clip]{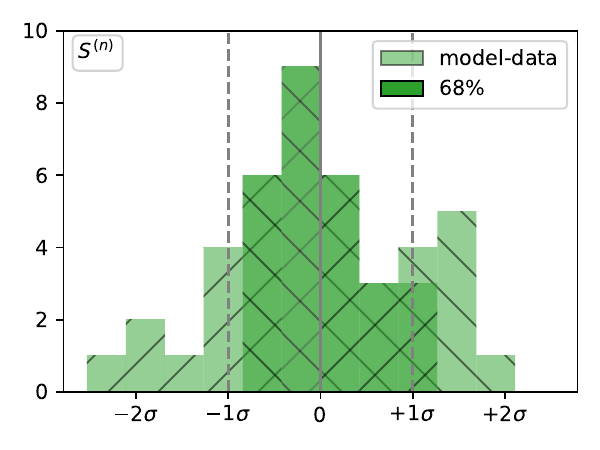}
    \caption{Same as Fig.~\ref{fig:Ca48_resid_dist} for $^{48}$Ti.}
    \label{fig:Ti48_resid_dist}
\end{figure}

\begin{figure}[p]
    \centering
    \includegraphics[width=0.99\linewidth, trim = 0 10 35 40,clip]{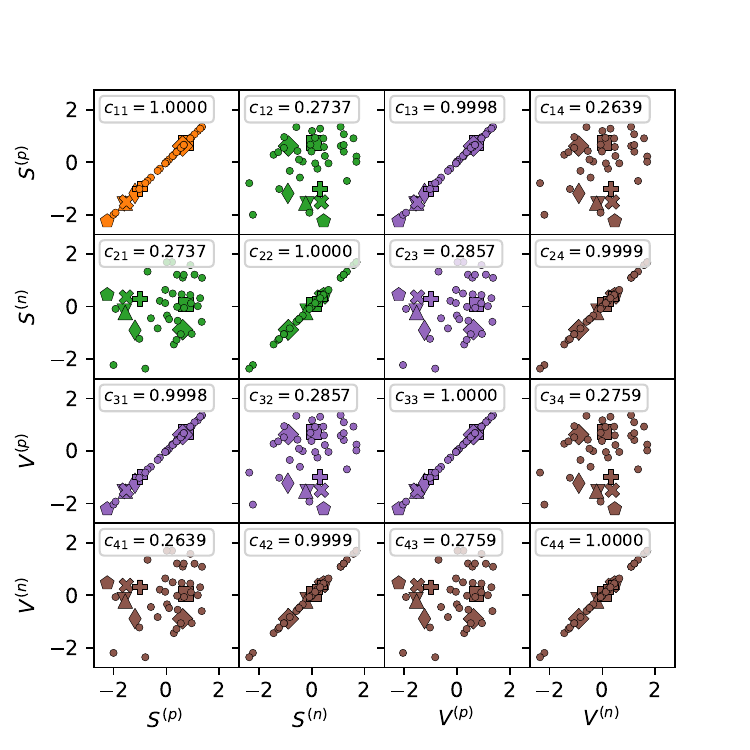}
    \caption{Correlations of the residuals for $^{48}$Ca normalized to the uncertainty based on the distribution as shown in Fig.~\ref{fig:Ca48_resid_dist}, using the same symbols as used in Fig.~1 of the main text. $c_{ij}$ refers to the correlation between the residuals of $I_i$ and $I_j$.}
    \label{fig:Ca48_resid_corr}
\end{figure}%

\begin{figure}[p]
    \centering
    \includegraphics[width=0.99\linewidth, trim = 0 10 35 40,clip]{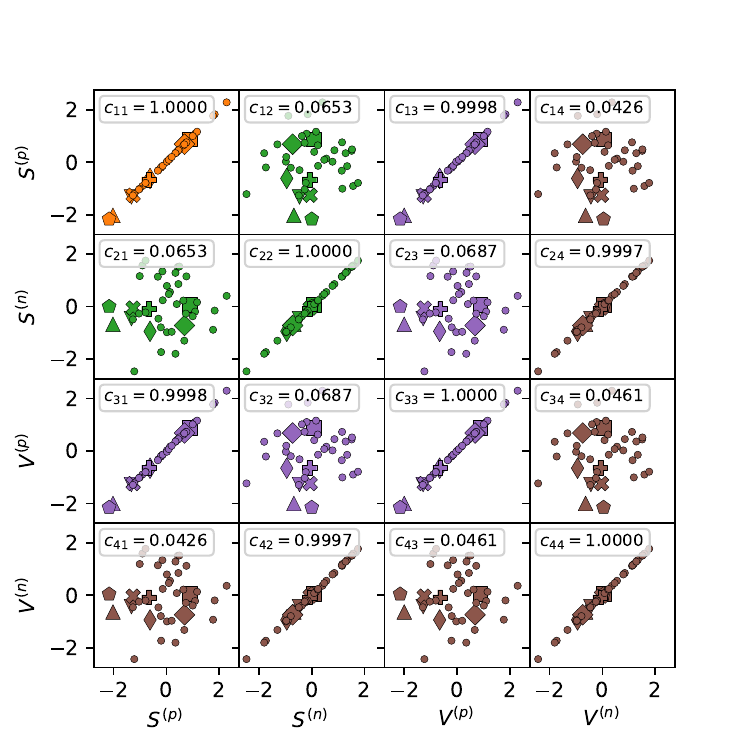}
    \caption{Same as Fig.~\ref{fig:Ca48_resid_corr} for $^{48}$Ti.}
    \label{fig:Ti48_resid_corr}
\end{figure}%

\begin{table}[H]
	\renewcommand{\arraystretch}{1.3} 
	\centering 
    \begin{tabular}{l l r r r r r r} 
	\toprule 
    & & & \multicolumn{1}{c}{This work/\cite{Noel:2024led}} & \multicolumn{2}{c}{References} \\ \hline
    \multirow{5}{*}{$^{27}$Al} & \multirow{2}{*}{$r_\text{ch}$} & \multirow{2}{*}{$[\text{fm}]$}  & \multirow{2}{*}{$3.063(31)$} & $3.035(2)$ & \cite{DeVries:1987atn} \\ 
    & & & & $3.0610(31)$ & \cite{Angeli:2013epw} \\
    & $r_\text{p}$ & $[\text{fm}]$ & $2.961(00)(32)$ & $2.925(7)$ & \cite{Qweak:2021ijt} \\
    & $r_\text{n}$ & $[\text{fm}]$ & $2.982(09)(33)$ & $2.89(12)$ & \cite{Qweak:2021ijt} \\
    & $r_\text{w}$ & $[\text{fm}]$ & $3.087(10)(32)$ & $3.00(15)$ & \cite{Qweak:2021ijt} \\ \hline
    \multirow{7}{*}{$^{48}$Ca} & \multirow{4}{*}{$r_\text{ch}$} & \multirow{4}{*}{$[\text{fm}]$} & \multirow{4}{*}{$3.475(10)$} & $3.451(9)$ & \cite{DeVries:1987atn} \\
    & & & & $3.4771(20)$ & \cite{Angeli:2013epw} \\
    & & & & $3.481$ & \cite{CREX:2022kgg} \\
    & & & & $3.48(3)$ & \cite{Hagen:2015yea} \\
    & $r_\text{p}$ & $[\text{fm}]$ & $3.403(00)(10)$ & $3.40(3)^*$ & \cite{Hagen:2015yea} \\
    & $r_\text{n}$ & $[\text{fm}]$ & $3.555(17)(11)$ & $[3.47,\,3.60]$ & \cite{Hagen:2015yea} \\
    & $r_\text{w}$ & $[\text{fm}]$ & $3.665(18)(11)$ & $[3.59,\,3.71]$ & \cite{Hagen:2015yea} \\ \hline
    \multirow{7}{*}{$^{48}$Ti} & \multirow{2}{*}{$r_\text{ch}$} & \multirow{2}{*}{$[\text{fm}]$} & \multirow{2}{*}{$3.5955(25)$} & $3.597(1)$ & \cite{DeVries:1987atn} \\
    & & & & $3.5921(17)$ & \cite{Angeli:2013epw} \\
    & $r_\text{p}$ & $[\text{fm}]$ & $3.515(00)(3)$ & \\
    & $r_\text{n}$ & $[\text{fm}]$ & $3.568(12)(3)$ & \\
    & $r_\text{w}$ & $[\text{fm}]$ & $3.665(12)(3)$ & \\
    & $r_\text{n}-r_\text{p}$ & $[\text{fm}]$ & $0.053(12)(4)$ & \\
    & $r_\text{w}-r_\text{ch}$ & $[\text{fm}]$ & $0.070(12)(4)$ & \\
    \bottomrule 
	\end{tabular} 

	\caption{Extension to Table~III in the main text, showing the absolute radii for $^{27}$Al and $^{48}$Ca (for $x\in\{\text{p},\text{n},\text{ch},\text{w}\}$ we define $r_x=\sqrt{\langle r^2 \rangle}_x$) and the analogous results for $^{48}$Ti, for which currently no experimental results from PVES exist. The values for the charge radius are taken from Ref.~\cite{Noel:2024led} and thus $r_\text{ch}=\sqrt{\expval{r^2}_\text{ch}^\text{ref}}$. The asterisk indicates that this value is not given directly in Ref.~\cite{Hagen:2015yea}, but can be inferred from the quantities provided.}
	\label{tab:r2}

\end{table}

\section*{Additional figures and tables}
\label{app:add}

\begin{figure}[H]
    \centering
    \includegraphics[width=0.99\linewidth,trim= 0 10 0 5,clip]{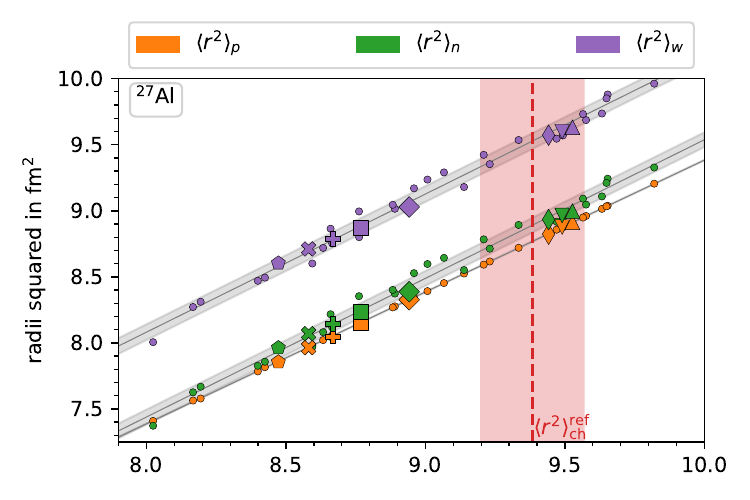}
    \includegraphics[width=0.99\linewidth,trim= 0 10 0 5,clip]{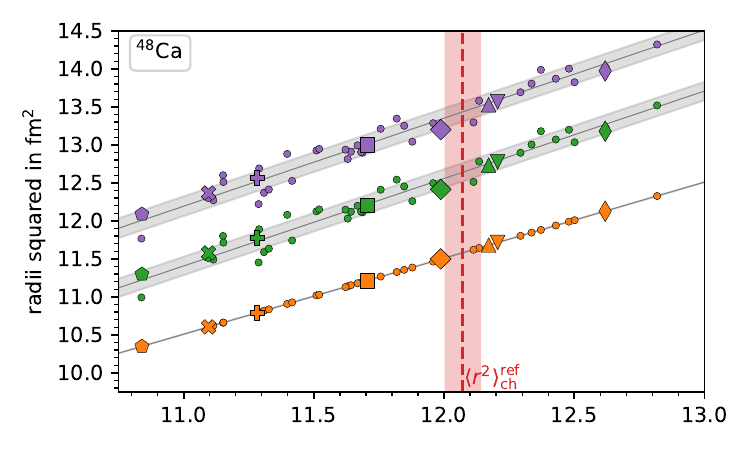}
    \includegraphics[width=0.99\linewidth,trim= 0 10 0 5,clip]{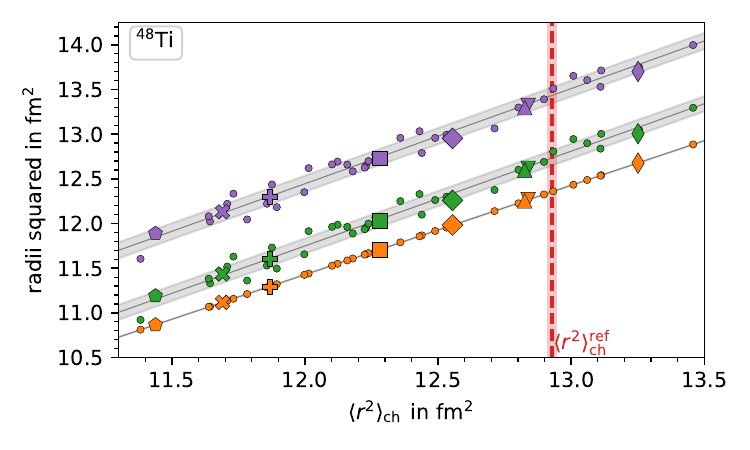}
    \caption{Correlations between $\expval{r^2}_\text{ch}$ and the point-proton, point-neutron, and weak radius squared, carried out in the same way as Fig.~1 of the main text.} 
    \label{fig:r2r2}
\end{figure}

In this section, we provide further material for the isotopes not discussed in full detail in the main text. First, Figs.~\ref{fig:Ca48_resid_dist}--\ref{fig:Ti48_resid_corr} show the residual distributions and their correlations for $^{48}$Ca and $^{48}$Ti. Table~\ref{tab:r2} extends Table~III in the main text by absolute radii for $^{27}$Al and $^{48}$Ca and the analogous results for $^{48}$Ti. The corresponding correlations, constructed in analogy to Fig.~1 in the main text, are shown in Fig.~\ref{fig:r2r2}.

\end{document}